\documentclass[journal]{IEEEtran}
\usepackage{amsmath,amsfonts}
\usepackage{algorithmic}
\usepackage{algorithm}
\usepackage{array}
\usepackage[caption=false,font=normalsize,labelfont=sf,textfont=sf]{subfig}
\usepackage{textcomp}
\usepackage{stfloats}
\usepackage{url}
\usepackage{verbatim}
\usepackage{graphicx}
\usepackage{cite}
\def\BibTeX{{\rm B\kern-.05em{\sc i\kern-.025em b}\kern-.08em
    T\kern-.1667em\lower.7ex\hbox{E}\kern-.125emX}}
    
\usepackage{acronym}
\usepackage{balance}

\usepackage{xcolor}

\begin{document}

\acrodef{ADC}[ADC]{Analog to Digital Converter}
\acrodef{ADEXP}[AdExp-I\&F]{Adaptive-Exponential Integrate and Fire}
\acrodef{ADM}[ADM]{Asynchronous Delta Modulator}
\acrodef{AER}[AER]{Address-Event Representation}
\acrodef{AEX}[AEX]{AER EXtension board}
\acrodef{AE}[AE]{Address-Event}
\acrodef{AFM}[AFM]{Atomic Force Microscope}
\acrodef{AGC}[AGC]{Automatic Gain Control}
\acrodef{AI}[AI]{Artificial Intelligence}
\acrodef{AMDA}[AMDA]{AER Motherboard with D/A converters}
\acrodef{ANN}[ANN]{Artificial Neural Network}
\acrodef{API}[API]{Application Programming Interface}
\acrodef{APMOM}[APMOM]{Alternate Polarity Metal On Metal}
\acrodef{ARM}[ARM]{Advanced RISC Machine}
\acrodef{ASIC}[ASIC]{Application Specific Integrated Circuit}
\acrodef{AdExp}[AdExp-IF]{Adaptive Exponential Integrate-and-Fire}
\acrodef{BCM}[BMC]{Bienenstock-Cooper-Munro}
\acrodef{BD}[BD]{Bundled Data}
\acrodef{BEOL}[BEOL]{Back-end of Line}
\acrodef{BG}[BG]{Bias Generator}
\acrodef{BPTT}[BPTT]{Backpropagation Through Time}
\acrodef{BMI}[BMI]{Brain-Machince Interface}
\acrodef{BTB}[BTB]{band-to-band tunnelling}
\acrodef{CAD}[CAD]{Computer Aided Design}
\acrodef{CAM}[CAM]{Content Addressable Memory}
\acrodef{CAVIAR}[CAVIAR]{Convolution AER Vision Architecture for Real-Time}
\acrodef{CA}[CA]{Cortical Automaton}
\acrodef{CCN}[CCN]{Cooperative and Competitive Network}
\acrodef{CDF}[CDF]{Cumulative Distribution Function}
\acrodef{CDR}[CDR]{Clock-Data Recovery}
\acrodef{CFC}[CFC]{Current to Frequency Converter}
\acrodef{CHP}[CHP]{Communicating Hardware Processes}
\acrodef{CMIM}[CMIM]{Metal-insulator-metal Capacitor}
\acrodef{CML}[CML]{Current Mode Logic}
\acrodef{CMOL}[CMOL]{Hybrid CMOS nanoelectronic circuits}
\acrodef{CMOS}[CMOS]{Complementary Metal-Oxide-Semiconductor}
\acrodef{CNN}[CCN]{Convolutional Neural Network}
\acrodef{COTS}[COTS]{Commercial Off-The-Shelf}
\acrodef{CPG}[CPG]{Central Pattern Generator}
\acrodef{CPLD}[CPLD]{Complex Programmable Logic Device}
\acrodef{CPU}[CPU]{Central Processing Unit}
\acrodef{CSM}[CSM]{Cortical State Machine}
\acrodef{CSP}[CSP]{Constraint Satisfaction Problem}
\acrodef{CTXCTL}[CTXCTL]{CortexControl}
\acrodef{CV}[CV]{Coefficient of Variation}
\acrodef{DAC}[DAC]{Digital to Analog Converter}
\acrodef{DAS}[DAS]{Dynamic Auditory Sensor}
\acrodef{DAVIS}[DAVIS]{Dynamic and Active Pixel Vision Sensor}
\acrodef{DBN}[DBN]{Deep Belief Network}
\acrodef{DFA}[DFA]{Deterministic Finite Automaton}
\acrodef{DIBL}[DIBL]{drain-induced-barrier-lowering}
\acrodef{DI}[DI]{delay insensitive}
\acrodef{DMA}[DMA]{Direct Memory Access}
\acrodef{DNF}[DNF]{Dynamic Neural Field}
\acrodef{DNN}[DNN]{Deep Neural Network}
\acrodef{DOF}[DOF]{Degrees of Freedom}
\acrodef{DPE}[DPE]{Dynamic Parameter Estimation}
\acrodef{DPI}[DPI]{Differential Pair Integrator}
\acrodef{DRAM}[DRAM]{Dynamic Random Access Memory}
\acrodef{DRRZ}[DR-RZ]{Dual-Rail Return-to-Zero}
\acrodef{DR}[DR]{Dual Rail}
\acrodef{DSP}[DSP]{Digital Signal Processor}
\acrodef{DVS}[DVS]{Dynamic Vision Sensor}
\acrodef{DYNAP}[DYNAP]{Dynamic Neuromorphic Asynchronous Processor}
\acrodef{EBL}[EBL]{Electron Beam Lithography}
\acrodef{EDVAC}[EDVAC]{Electronic Discrete Variable Automatic Computer}
\acrodef{EEG}[EEG]{electroencephalography}
\acrodef{EIN}[EIN]{Excitatory-Inhibitory Network}
\acrodef{EM}[EM]{Expectation Maximization}
\acrodef{EPSC}[EPSC]{Excitatory Post-Synaptic Current}
\acrodef{EPSP}[EPSP]{Excitatory Post-Synaptic Potential}
\acrodef{EZ}[EZ]{Epileptogenic Zone}
\acrodef{FDSOI}[FDSOI]{Fully-Depleted Silicon on Insulator}
\acrodef{FET}[FET]{Field-Effect Transistor}
\acrodef{FFT}[FFT]{Fast Fourier Transform}
\acrodef{FI}[F-I]{Frequency-Current}
\acrodef{FPGA}[FPGA]{Field Programmable Gate Array}
\acrodef{FR}[FR]{Fast Ripple}
\acrodef{FSA}[FSA]{Finite State Automaton}
\acrodef{FSM}[FSM]{Finite State Machine}
\acrodef{GIDL}[GIDL]{gate-induced-drain-leakage}
\acrodef{GOPS}[GOPS]{Giga-Operations per Second}
\acrodef{GPU}[GPU]{Graphical Processing Unit}
\acrodef{GUI}[GUI]{Graphical User Interface}
\acrodef{HAL}[HAL]{Hardware Abstraction Layer}
\acrodef{HFO}[HFO]{High Frequency Oscillation}
\acrodef{HH}[H\&H]{Hodgkin \& Huxley}
\acrodef{HMM}[HMM]{Hidden Markov Model}
\acrodef{HRS}[HRS]{High-Resistive State}
\acrodef{HR}[HR]{Human Readable}
\acrodef{HSE}[HSE]{Handshaking Expansion}
\acrodef{HW}[HW]{Hardware}
\acrodef{ICT}[ICT]{Information and Communication Technology}
\acrodef{IC}[IC]{Integrated Circuit}
\acrodef{IEEG}[iEEG]{intracranial electroencephalography}
\acrodef{IF2DWTA}[IF2DWTA]{Integrate \& Fire 2--Dimensional WTA}
\acrodef{IFSLWTA}[IFSLWTA]{Integrate \& Fire Stop Learning WTA}
\acrodef{IF}[I\&F]{Integrate-and-Fire}
\acrodef{IMU}[IMU]{Inertial Measurement Unit}
\acrodef{INCF}[INCF]{International Neuroinformatics Coordinating Facility}
\acrodef{INI}[INI]{Institute of Neuroinformatics}
\acrodef{IO}[I/O]{Input/Output}
\acrodef{IPSC}[IPSC]{Inhibitory Post-Synaptic Current}
\acrodef{IPSP}[IPSP]{Inhibitory Post-Synaptic Potential}
\acrodef{IP}[IP]{Intrinsic Plasticity}
\acrodef{ISI}[ISI]{Inter-Spike Interval}
\acrodef{IoT}[IoT]{Internet of Things}
\acrodef{JFLAP}[JFLAP]{Java - Formal Languages and Automata Package}
\acrodef{LEDR}[LEDR]{Level-Encoded Dual-Rail}
\acrodef{LIF}[LIF]{Leaky Integrate and Fire}
\acrodef{LFP}[LFP]{Local Field Potential}
\acrodef{LLC}[LLC]{Low Leakage Cell}
\acrodef{LNA}[LNA]{Low-Noise Amplifier}
\acrodef{LPF}[LPF]{Low Pass Filter}
\acrodef{LRS}[LRS]{Low-Resistive State}
\acrodef{LSM}[LSM]{Liquid State Machine}
\acrodef{LTD}[LTD]{Long Term Depression}
\acrodef{LTI}[LTI]{Linear Time-Invariant}
\acrodef{LTP}[LTP]{Long Term Potentiation}
\acrodef{LTU}[LTU]{Linear Threshold Unit}
\acrodef{LUT}[LUT]{Look-Up Table}
\acrodef{LVDS}[LVDS]{Low Voltage Differential Signaling}
\acrodef{MCMC}[MCMC]{Markov-Chain Monte Carlo}
\acrodef{MEMS}[MEMS]{Micro Electro Mechanical System}
\acrodef{MFR}[MFR]{Mean Firing Rate}
\acrodef{MIM}[MIM]{Metal Insulator Metal}
\acrodef{MLP}[MLP]{Multilayer Perceptron}
\acrodef{MOSCAP}[MOSCAP]{Metal Oxide Semiconductor Capacitor}
\acrodef{MOSFET}[MOSFET]{Metal Oxide Semiconductor Field-Effect Transistor}
\acrodef{MOS}[MOS]{Metal Oxide Semiconductor}
\acrodef{MRI}[MRI]{Magnetic Resonance Imaging}
\acrodef{NDFSM}[NDFSM]{Non-deterministic Finite State Machine} 
\acrodef{ND}[ND]{Noise-Driven}
\acrodef{NEF}[NEF]{Neural Engineering Framework}
\acrodef{NHML}[NHML]{Neuromorphic Hardware Mark-up Language}
\acrodef{NIL}[NIL]{Nano-Imprint Lithography}
\acrodef{NMDA}[NMDA]{N-Methyl-D-Aspartate}
\acrodef{NME}[NE]{Neuromorphic Engineering}
\acrodef{NN}[NN]{Neural Network}
\acrodef{NRZ}[NRZ]{Non-Return-to-Zero}
\acrodef{NSM}[NSM]{Neural State Machine}
\acrodef{OR}[OR]{Operating Room}
\acrodef{OTA}[OTA]{Operational Transconductance Amplifier}
\acrodef{PCA}[PCA]{Principle Component Analysis}
\acrodef{PCB}[PCB]{Printed Circuit Board}
\acrodef{PCHB}[PCHB]{Pre-Charge Half-Buffer}
\acrodef{PCM}[PCM]{Phase Change Memory}
\acrodef{PDF}[PDF]{Probability Distribution Function}
\acrodef{PE}[PE]{Phase Encoding}
\acrodef{PFA}[PFA]{Probabilistic Finite Automaton}
\acrodef{PFC}[PFC]{prefrontal cortex}
\acrodef{PFM}[PFM]{Pulse Frequency Modulation}
\acrodef{PR}[PR]{Production Rule}
\acrodef{PSC}[PSC]{Post-Synaptic Current}
\acrodef{PSP}[PSP]{Post-Synaptic Potential}
\acrodef{PSTH}[PSTH]{Peri-Stimulus Time Histogram}
\acrodef{QDI}[QDI]{Quasi Delay Insensitive}
\acrodef{RAM}[RAM]{Random Access Memory}
\acrodef{RA}[RA]{Resected Area}
\acrodef{RDF}[RDF]{random dopant fluctuation}
\acrodef{RELU}[ReLu]{Rectified Linear Unit}
\acrodef{RLS}[RLS]{Recursive Least-Squares}
\acrodef{RMSE}[RMSE]{Root Mean Squared-Error}
\acrodef{RMS}[RMS]{Root Mean Squared}
\acrodef{RNN}[RNN]{Recurrent Neural Network}
\acrodef{ROLLS}[ROLLS]{Reconfigurable On-Line Learning Spiking}
\acrodef{RRAM}[RRAM]{Resistive Random Access Memory}
\acrodef{ReRAM}[ReRAM]{Resistive Random Access Memory}
\acrodef{R}[R]{Ripples}
\acrodef{SAC}[SAC]{Selective Attention Chip}
\acrodef{SAT}[SAT]{Boolean Satisfiability Problem}
\acrodef{SCX}[SCX]{Silicon CorteX}
\acrodef{SD}[SD]{Signal-Driven}
\acrodef{SDSP}[SDSP]{Spike Driven Synaptic Plasticity}
\acrodef{SEM}[SEM]{Spike-based Expectation Maximization}
\acrodef{SGD}[SGD]{Stochastic Gradient Descent}
\acrodef{SLAM}[SLAM]{Simultaneous Localization and Mapping}
\acrodef{SORN}[SORN]{Self-Organizing Recurrent Network}
\acrodef{SNN}[SNN]{Spiking Neural Network}
\acrodef{SNR}[SNR]{Signal to Noise Ratio}
\acrodef{SOC}[SOC]{System-On-Chip}
\acrodef{SOI}[SOI]{Silicon on Insulator}
\acrodef{SOZ}[SOZ]{Seizure Onset Zone}
\acrodef{SP}[SP]{Separation Property}
\acrodef{SRAM}[SRAM]{Static Random Access Memory}
\acrodef{SRNN}[SRNN]{Spiking Recurrent Neural Network}
\acrodef{STDP}[STDP]{Spike-Timing Dependent Plasticity}
\acrodef{STD}[STD]{Short-Term Depression}
\acrodef{STP}[STP]{Short-Term Plasticity}
\acrodef{STT-MRAM}[STT-MRAM]{Spin-Transfer Torque Magnetic Random Access Memory}
\acrodef{STT}[STT]{Spin-Transfer Torque}
\acrodef{SW}[SW]{Software}
\acrodef{TCAM}[TCAM]{Ternary Content-Addressable Memory}
\acrodef{TFT}[TFT]{Thin Film Transistor}
\acrodef{TLE}[TLE]{Temporal Lobe Epilepsy}
\acrodef{USB}[USB]{Universal Serial Bus}
\acrodef{VHDL}[VHDL]{VHSIC Hardware Description Language}
\acrodef{VLSI}[VLSI]{Very Large Scale Integration}
\acrodef{VOR}[VOR]{Vestibulo-Ocular Reflex}
\acrodef{WCST}[WCST]{Wisconsin Card Sorting Test}
\acrodef{WTA}[WTA]{Winner-Take-All}
\acrodef{XML}[XML]{eXtensible Mark-up Language}
\acrodef{divmod3}[DIVMOD3]{divisibility of a number by three}
\acrodef{hWTA}[hWTA]{hard Winner-Take-All}
\acrodef{sWTA}[sWTA]{soft Winner-Take-All}

\acrodef{SOSN}[MEMSORN]{Memristive Self-organizing Spiking Recurrent Neural Network}

\title{
Scaling Limits of Memristor-Based Routers for Asynchronous Neuromorphic Systems
}
\author{Junren Chen, Siyao Yang, Huaqiang Wu,~\IEEEmembership{Senior Member,~IEEE}, Giacomo Indiveri,~\IEEEmembership{Senior Member,~IEEE}, \\Melika Payvand,~\IEEEmembership{Member, IEEE}
\thanks{This work won the Best Paper Award at the 2023 30th IEEE International Conference on Electronics, Circuits and Systems (ICECS).

Junren Chen, Giacomo Indiveri and Melika Payvand are with the Institute of Neuroinformatics, University of Zurich and ETH Zurich, 8057 Zurich, Switzerland (e-mail: junren@ini.uzh.ch). 

Siyao Yang and Huaqiang Wu
are with the School of Integrated Circuits, Tsinghua University, 100084 Beijing, China.

Manuscript received MONTH DD, 2023; revised MONTH DD, YEAR.
}
}

\markboth{IEEE Transactions on Circuits and Systems II: Express Briefs, VOL. X, NO. X, December 2023}%
{J. Chen \MakeLowercase{\textit{et al.}}: Scaling Limits of Memristor-Based Routers for Asynchronous Neuromorphic Systems}


\maketitle

\begin{abstract}
Multi-core neuromorphic systems typically use on-chip routers to transmit spikes among cores. These routers require significant memory resources and consume a large part of the overall system's energy budget.
A promising alternative approach to using standard CMOS and SRAM-based routers is to exploit the features of memristive crossbar arrays and use them as programmable switch-matrices that route spikes.
However, the scaling of these crossbar arrays presents physical challenges, such as ``IR drop'' on the metal lines due to the parasitic resistance, and leakage current accumulation on multiple active memristors in their ``off'' state. 
While reliability challenges of this type have been extensively studied in synchronous systems for compute-in-memory matrix-vector multiplication (MVM) accelerators and storage class memory, little effort has been devoted so far to characterizing the scaling limits of memristor-based crossbar routers. Here, we study the challenges of memristive crossbar arrays, when used as routing channels to transmit spikes in asynchronous Spiking Neural Network (SNN) hardware. We validate our analytical findings with experimental results obtained from a 4K-ReRAM chip which demonstrates its functionality as a routing crossbar. We determine the functionality bounds on the routing due to the IR drop and leak problem, based on theoretical modeling, circuit simulations for a 22\,nm FDSOI technology, and experimental measurements. 
This work highlights the limitations of this approach and provides useful guidelines for engineering the memristor device properties in memristive crossbar routers for multi-core asynchronous neuromorphic systems. 

\end{abstract}


\section{Introduction}
\label{sec:introduction}

\begin{figure}[!t]
\centerline{\includegraphics[scale=0.4, trim={0cm 0.cm 0cm 0.1cm},clip]{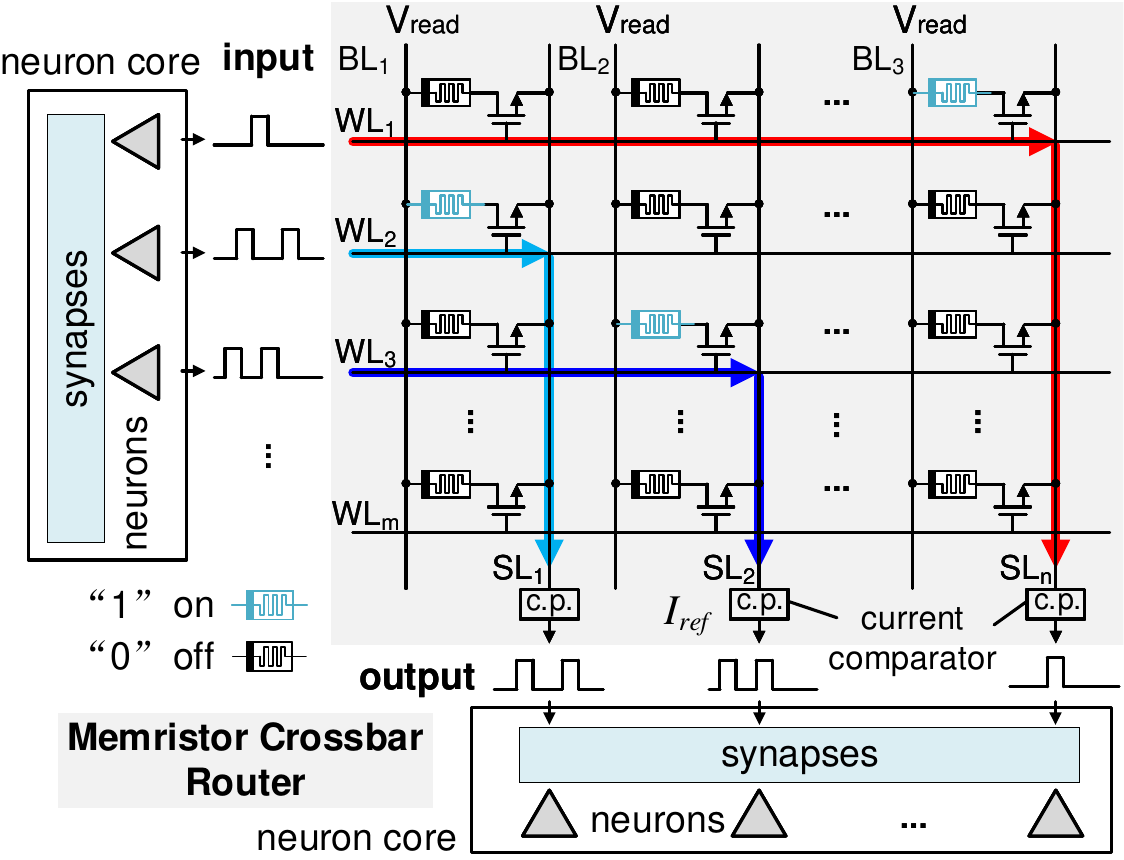}}
\caption{Schematic diagram of a memristor-based switch-matrix for transmitting voltage pulses (spikes) across neural cores, using 1-transistor-1-memristor (1T1R) crossbars. One column represents a single routing channel. $V_{read}$ is applied to read the cells at the arrival of input pulses on WLs. If the current on SL is greater than a reference ($I_{ref}$), the current comparator (c.p.) generates an output pulse, otherwise, the output remains at zero voltage. Thus, the ``on'' state, i.e., Low Resistance State (LRS), of the memristor enables the propagation of the input spikes,  while the “off” state, i.e., High Resistance State (HRS), blocks them.
}
\label{fig:intro}
\end{figure}

\IEEEPARstart{W}{hile} memristive devices have been often used in neuromorphic systems to emulate synaptic functions~\cite{Indiveri_2013,yao2017face}, their non-volatile memory features can also be exploited to configure and re-program the connectivity between digital logic elements~\cite{xia2009memristor,ISSCC_RRAM_FPGA,2020RRAM_FPGA}. 
This memristor-based routing approach can also be used in multi-core \ac{SNN} neuromorphic processors, in which neurons communicate with each other via all-or-none spike events~\cite{dalgaty_etal2021_mosaic,Point2Point_TCAS_II}. Memristive crossbar routers act as distributed switch matrices which allow the digital spikes to be transmitted or get blocked, thus resulting in programming and storing the connectivity of the \ac{SNN} between multiple processing cores (Fig.~\ref{fig:intro}). 
This approach can be particularly advantageous for reducing static power consumption, thanks to the very low static power consumption of resistive memory, and to their non-volatility.
As static power consumption accounts for a large proportion of the power budget in CMOS memories implemented in advanced technologies nodes~\cite{2003LeakPower}, memristive routers represent a promising alternative for building multi-core \ac{SNN} processors on stringent power budgets (e.g. for  edge-computing applications).

To scale the fan-in/out of the neurons while keeping the memory resources at minimum, it is however important to leverage the temporal sparsity of \acp{SNN} so that routing channels can be used as shared resources for multiple neurons. A theoretical reliability study of the trade-off between fan-in/out of an \ac{SNN} node and transmission collision probability in a routing channel as a function of array size has already been reported in~\cite{chen2022reliability}. This study has also characterized the reliability of signal transmission with respect to the on/off ratio ($R_{on}/R_{off}$) of individual memristive devices. 
In this work, we extend this study with both further analysis on the scaling of memristive routing arrays, and by performing experimental measurements on a fabricated 4K-ReRAM chip.
Specifically we first experimentally demonstrate the ability of a memristor crossbar router to transmit spikes, and characterize its scaling reliability with respect to the leakage current accumulation from ``off'' cells. Next, we present a quantitative analysis, using both theoretical considerations and circuit simulations, on the issue of scaling due to  ``IR drop'' as a result of the line parasitics. Finally, we evaluate the influence of leakage current from MOSFETs when scaling down to 22\,nm technology. We conclude by providing guidelines and specifications on the memristor properties required for enabling the scaling of neuromorphic processing systems that use memristor-based routing schemes. 

It is worth noting that while the memristor crossbar router in this work is specific to SNN hardware, the routing approach and principles discussed may find broader applications.

\begin{figure*}[!t]
\centerline{\includegraphics[scale=0.42, trim={0.cm 0.25cm 0cm 0.1cm},clip]{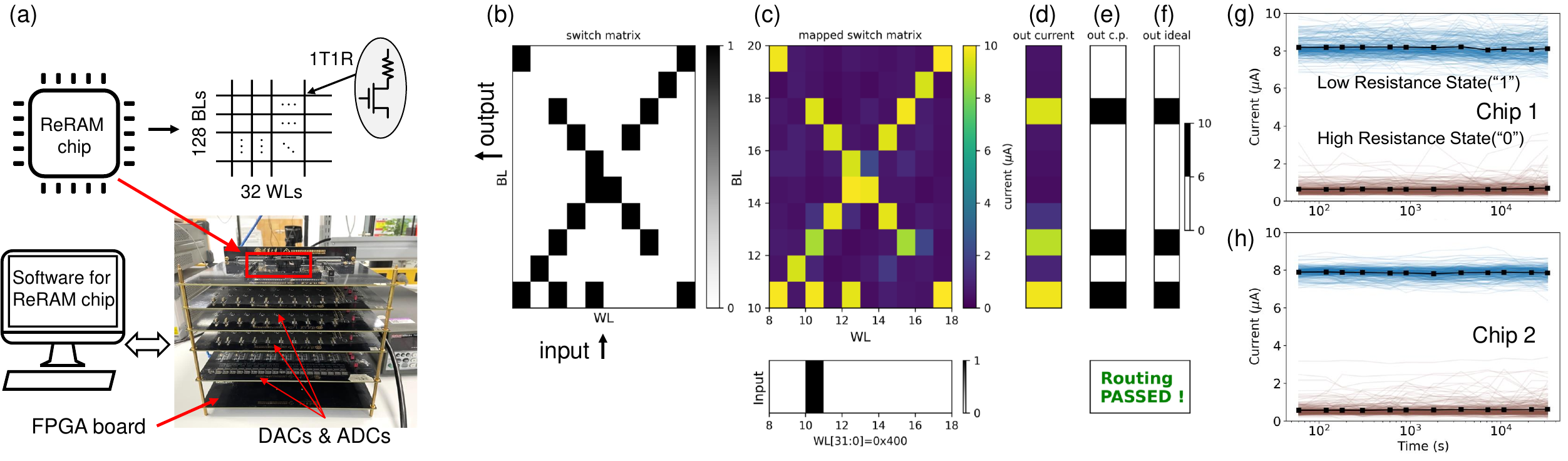}}
\caption{Experimental demonstration of routing spikes on a memristor crossbar chip. (a) The measurement system of a 4K-cell ReRAM chip. 
(b) An arbitrary switch matrix pattern. The black squares indicate the cross points that connect the corresponding inputs and outputs. (c) The mapped switch matrix on the hardware crossbar array. The colors show the currents from each cell (read at $0.2\,V$). (d) Read out currents at the arrival of an input spike. Input spike on WL$_{10}$ is multicast to three receivers (e.g. neurons). (e) The output currents quantified by a threshold (6\,$\mu A$), which simulates the function of a current comparator (c.p.). (f) The ideal correct output pulses corresponding to applying the input on the switch matrix in (b). If (e) matches (f), the routing is successful (``passed''), otherwise the routing is ``failed'' due to transmission errors. (g)\&(h)  Retention measurements of binary ReRAM cells of two different chips at room temperature, demonstrating the reliability of the memory devices for binary operation and chip-to-chip variability. Black lines show the mean values (for 256 cells/state).}

\label{fig:RoutingDemo}
\end{figure*}

\section{Experimental measurements}
\label{sec:experiment}

\subsection{Memristor Crossbar Router on a ReRAM Chip}

The experimental measurements were made using non-volatile 1T1R HfO$_x$-based \ac{ReRAM} chips integrating $32\times128$ (4K) cells.
An FPGA board in the measurement system supports over two hundred pins of the ReRAM chip operating in parallel, which  provides very high flexibility and test capacity. For instance, 32 WLs can be operated individually and in parallel. The system supports parallel read out. It can flexibly implement unicast, multicast and broadcast communication by programming one, multiple or all cells on the same input Word Line (WL) to ``on'' state, respectively. This feature is key for enabling the use of a chip as crossbar router demonstrator. 
Figure~\ref{fig:RoutingDemo} shows the experimental set up and demonstrates routing spikes through the ReRAM chip. Note that the presentation of the crossbar (WL, BL locations) in Fig.~\ref{fig:RoutingDemo} is rotated {90\textdegree} compared to Fig.~\ref{fig:intro}. The input/output directions are labeled accordingly for the clarification. An arbitrary switch matrix (b) is mapped on the array (c), and an input spike on WL$_{10}$ is  sent to three receivers. Since the c.p. output (e) matches the ideal output (f), routing is successful.
The retention of binary cells of two different chips (g, h) show the reliability of ReRAM cells.  

\subsection{Undesired Output Pulses from ``Off'' Cells}
\label{sec:underired_output}
\begin{figure}[!t]
\centerline{\includegraphics[scale=0.48, trim={0 0.05cm 0 0.1cm},clip]{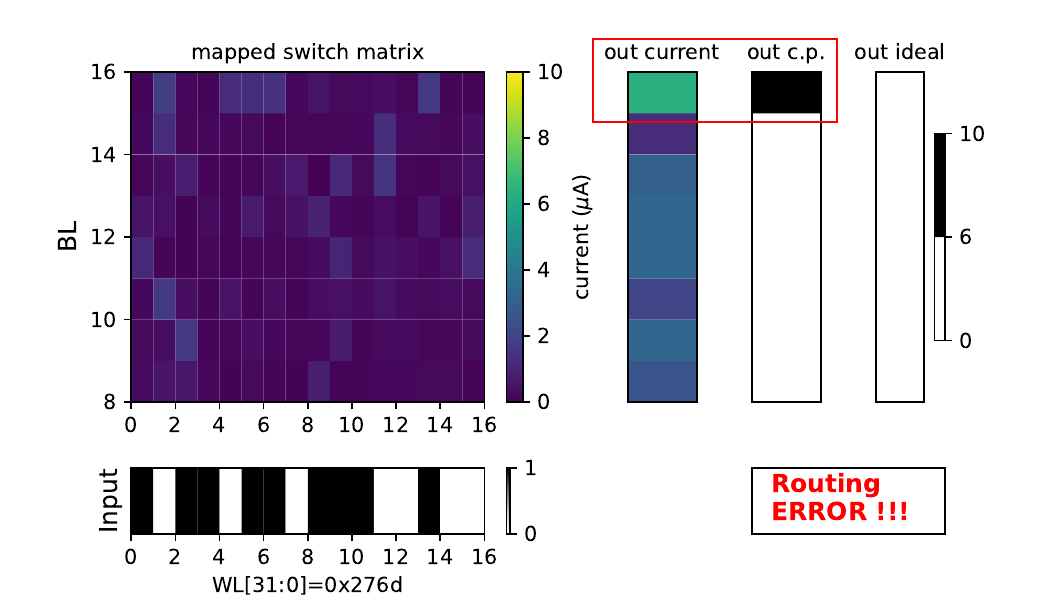}}
\caption{A measurement showing that when nine simultaneous input pulses arrive and all cells are in ``off'' state, an undesired error output pulse is generated.}
\label{fig:error}
\end{figure}

The leakage from the active WLs of the crossbar array with ``off'' cells can pose a reliability issue. When multiple input pulses arrive at the same time, an undesired error pulse will be generated, if the accumulated current on a SL from ``off'' cells is higher than the threshold of c.p. ($I_{ref}$). 
This issue indicates a minimum on/off ratio requirement of memristors to avoid transmission errors, as determined also in our previous work~\cite{chen2022reliability}. Here, we experimentally measure the occurrence of these errors (see Fig.~\ref{fig:error}) for a case in which nine simultaneous input pulses arrive and all cells are programmed to ``off'' state. In this experiment, on the bit-line BL$_{15}$, the accumulated ``off'' leakage currents is higher than the threshold (6\,$\mu A$), so an output routing pulse is generated, even though no valid routing path exists.

\begin{figure}[!t]
\centerline{\includegraphics[scale=0.18, trim={0.cm 0.cm 0.1cm 0.cm},clip]{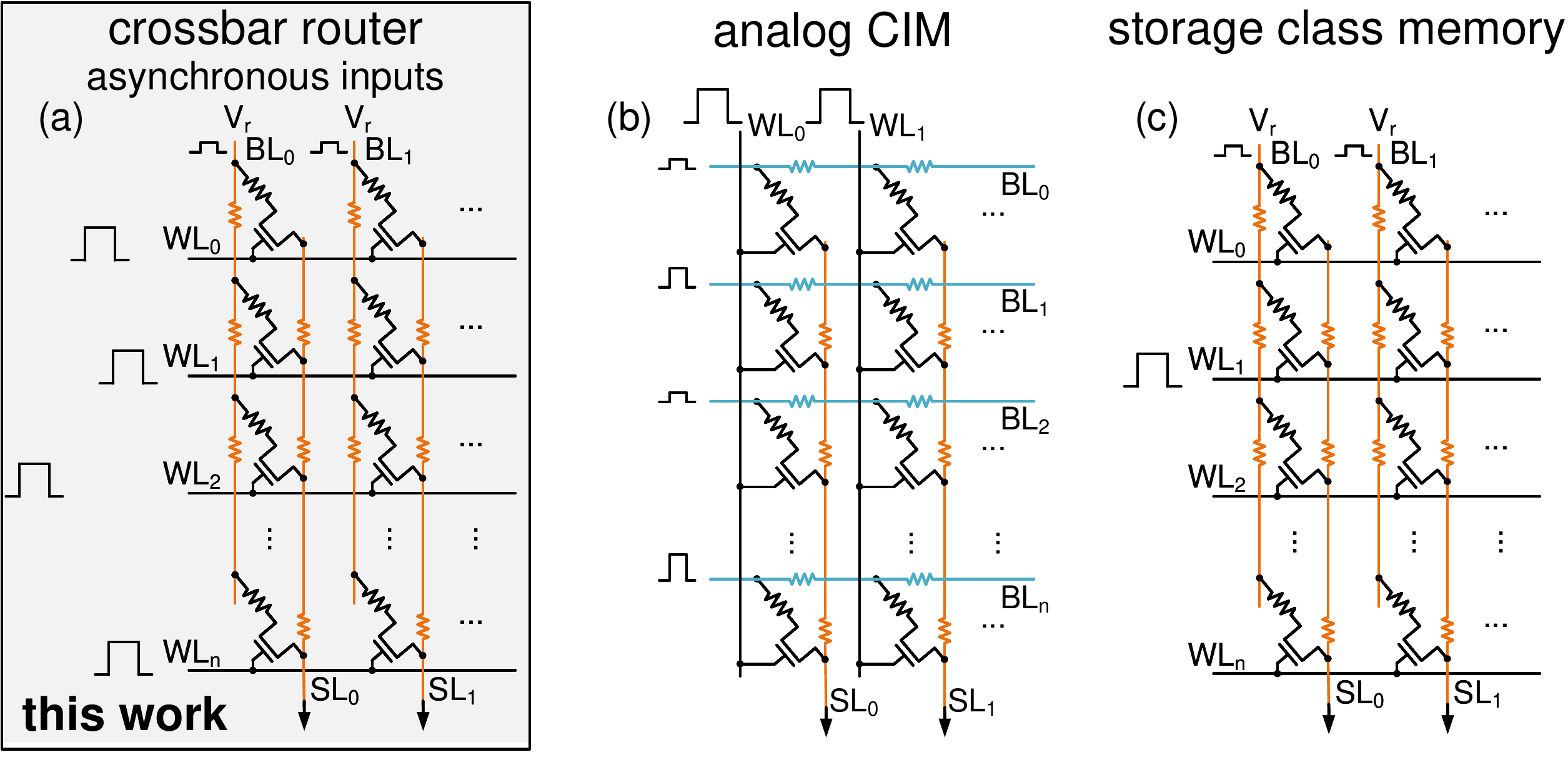}}
\caption{IR drop pathways for different applications. 
(a) Crossbar router diagram: transmitting asynchronous voltage pulses. (b) Compute-in-memory (CIM) MVM accelerators in synchronous system: fully parallel computing. (c) Storage class memory: only one WL is activated in a read operation in synchronous/asynchronous system. 
}
\vspace{-0.1cm}
\label{fig:IR_schematic}
\end{figure}

\section{Scaling analysis}
\label{sec:scaling}
In addition to the leak accumulation problem, the IR drop on the metal lines poses a challenge for scaling memristor crossbars. In the crossbar router case, it decreases the read margin between ``on'' and ``off'' states.

\subsection{IR Drop Analysis}
\label{sec:IRdrop_router} 
Figure~\ref{fig:IR_schematic} shows the IR drop paths on 1T1R array with different WL, Bit Line (BL) and Source Line (SL) arrangements. The input patterns differentiate between different applications. IR drop reliability issues on CIM MVM accelerators and storage class memory have been extensively studied in previous works, e.g.~\cite{Liao2020, Qin2022, 2013XB_scaling}. 
For our crossbar router use case (Fig.~\ref{fig:IR_schematic} (a)), we evaluate the routing error probability caused by current accumulation from ``off'' cells as a function of the network/router scaling. 

\begin{figure}[!t]
\centering
\includegraphics[scale=0.26, trim={0cm 0.4cm 0cm 0.1cm},clip]{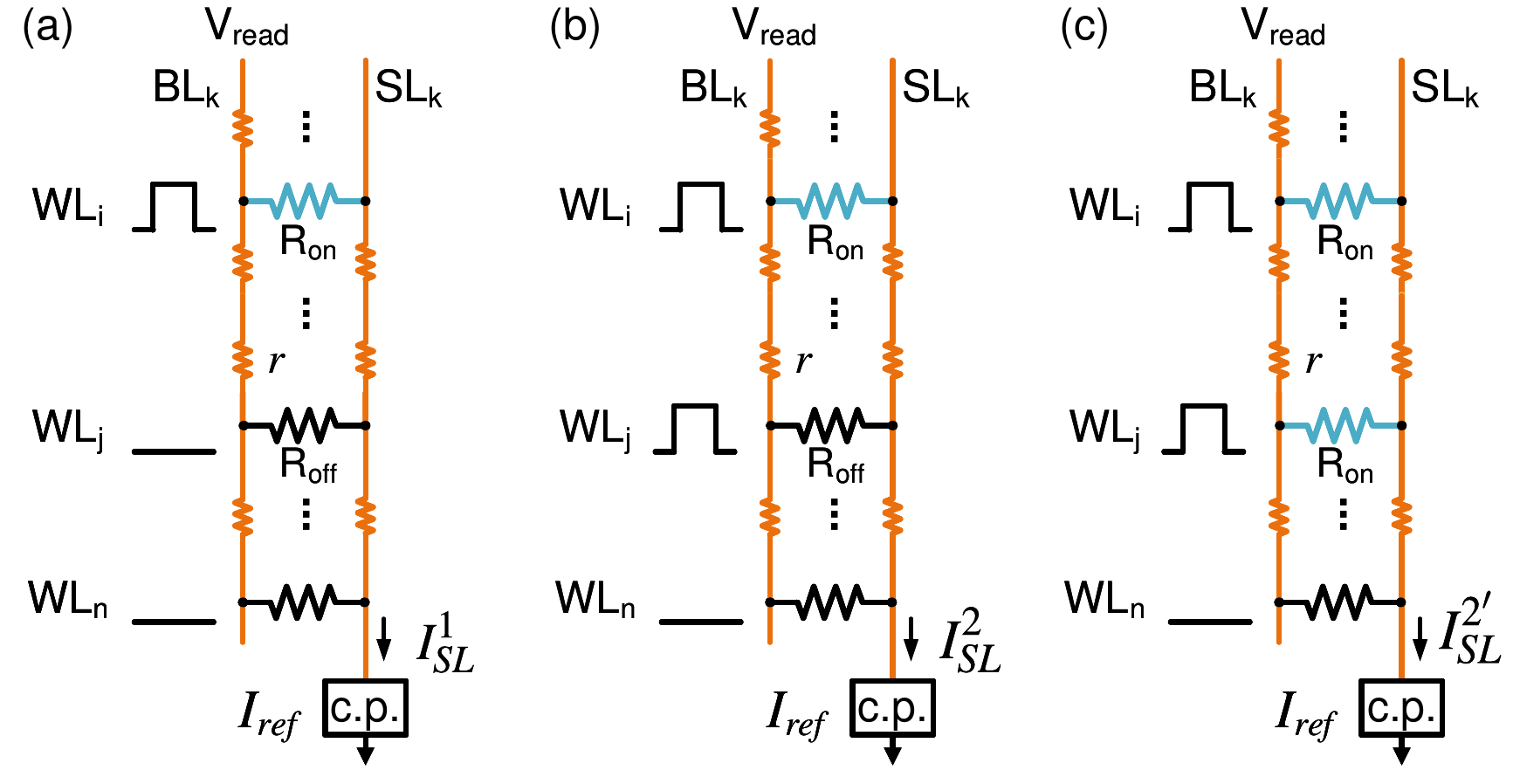}
\caption{IR drop model of one routing channel. $r$ is the unit line resistance. (a) Only one $R_{on}$ is switched on. (b) One $R_{on}$ and one $R_{off}$ are switched on at the same time. (c) Two $R_{on}$ are switched on at the same time.}
\label{fig:IRdrop_model}
\vspace{-0.2cm}
\end{figure}

Due to the IR drop, the read voltage applied to the memory cells down the line will be lower than the one applied, resulting in lower output currents, and hence resulting in a possible output error in the corresponding routing channel. This is illustrated in Fig.~\ref{fig:IRdrop_model}. 
For example, when reading a cell which stores a ``1'', and expecting to transmit the input pulse to the corresponding output. There will be an error if the IR drop leads to $I_{SL} < I_{ref}$, as no output voltage pulse will be generated. Since the IR drop only has the effect of decreasing the values of $I_{SL}$, it will not cause errors for ``0'' cases, so we restrict our analysis to the reading of ``1'' cases.

In the example shown in Fig.~\ref{fig:IRdrop_model} (a), only one spike arrives to the channel on one $R_{on}$ (WL$_{i}$). Assuming $n > j > i$,
\begin{equation}  \label{eq:ISL_1}
    \small
    \begin{aligned}
    I_{SL}^1 &= \frac{V_{read}}{ir + R_{on} + (n-i)r} = \frac{V_{read}}{R_{on} + nr}
    \end{aligned}
\end{equation}
where $r$ is the line resistance between two adjacent cells. $R_{on}$ is the resistance of a ``on'' cell.

In the example of Fig.~\ref{fig:IRdrop_model} (b), when two spikes arrive: WL$_i$ is switched on, while another $R_{off}$ (WL$_j$) is on,
\begin{equation}
\label{eq:ISL_2_1}
\small
    \begin{aligned}
    I_{SL}^2 &= \frac{V_{read}}{ir + [R_{on} + (j-i)r] // [(j-i)r + R_{off}] + (n-j)r}
    \end{aligned}
\end{equation}

Since $[R_{on} + (j-i)r] // [(j-i)r + R_{off}] < R_{on} + (j-i)r$, we can derive:
\begin{equation}  \label{eq:ISL_2}
\small
    \begin{aligned}
    I_{SL}^2 &> \frac{V_{read}}{ir + R_{on} + (j-i)r + (n-j)r} = \frac{V_{read}}{R_{on} + nr}  = I_{SL}^1
    \end{aligned}
\end{equation}

We get $I_{SL}^2 > I_{SL}^1$. Accordingly, $I_{SL}^n > ... > I_{SL}^3 > I_{SL}^2 > I_{SL}^1$, $n$ means $n$ cells are simultaneously activated. If $R_{off}$ in eq.~\ref{eq:ISL_2_1} is replaced with $R_{on}$, i.e. case Fig.~\ref{fig:IRdrop_model} (c), we can get the same result that $I_{SL}^{n'} > ... > I_{SL}^{3'} > I_{SL}^{2'} > I_{SL}^{1}$. It means that any additional cells (either $R_{off}$ or $R_{on}$) in the same channel are switched on with one ``on" path, the resultant output current on the SL is higher than $I_{SL}^1$. So, as long as $I_{SL}^1$ is able to produce a correct routing output, i.e. the condition $I_{SL}^1 > I_{ref}$ holds, any additional simultaneous inputs will not lead to errors, no matter how big $r$ is. 

Thus, for what concerns the input patterns, simultaneous inputs do not cause routing errors due to IR drop.
The impact of IR drop on the functionality of the crossbar router that we need to assess is the margin to separate a ``on'' cell and the accumulation of multiple ``off'' ones. 

Defining the on/off (or in our case off/on) ratio of fabricated memristors as $k = R_{off}/R_{on} > 1$.
The physical sensing circuit at the end of SL or BL has to always take into account also the resistance of the metal line ($nr$). So, the real on/off ratio of memristors relevant for the sensing circuit becomes $k' = (R_{off} + nr)/(R_{on} + nr)$.
It should be noted that practically, $ k'$ is the ratio of the reference current of sensing circuit to the current of ``off'' cells, not exactly ``on'' cell to ``off'' cell when device-to-device variation is factored in. 

The relationship between $k$ and $k'$ is
\begin{equation}  \label{eq:k_k'}
\small
    \begin{aligned}
     k' &= \frac{k R_{on} + nr}{R_{on} + nr} = \frac{(k-1)R_{on} + R_{on} + nr}{R_{on} + nr} = \frac{k-1}{1 + nr/R_{on}} + 1
    \end{aligned}
\end{equation}

Since $1 + nr/R_{on} > 1$, $\frac{k-1}{1 + nr/R_{on}} < k-1$, $k'< k$. 
As the router size scales up and the metal pitch of the fabrication technology scales down, IR drop increases and on/off ratio decreases. For example, if $nr=R_{on}$, the on/off ratio to sensing circuit is $ k' = k/2 + 1/2$. The sensing window (SW) is halved.  
IR drop lowers SW, which can increase error probability caused by accumulated currents from ``off'' cells.

\begin{figure}[!t]
\centering
\centerline{\includegraphics[scale=0.32, trim={0.5 0.15cm 0 0.4cm},clip]{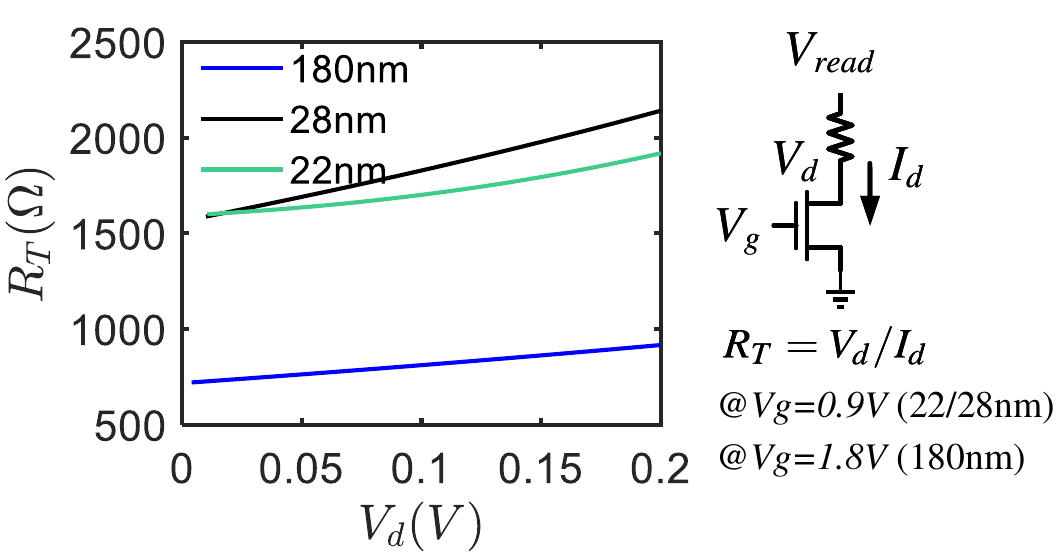}}
\caption{Resistance of transistor in 1T1R cells in different technology nodes. Simulation based on PDK from foundries. Transistor size is chosen with proper driving capacity for programming memristors ($>150uA$).}
\label{fig:Rt}
\end{figure}

\subsection{Influence of Resistance in Transistors}
\label{sec:Rt_router}
In practice, to engineer the proper resistance range of memristors and their scaling ability (limited by IR drop and on/off ratio requirement) for routing application in the 1T1R scheme of Fig.~\ref{fig:intro}, the resistance of transistors in routing mode (when reading memristors) should be taken into account. 
We can therefore reformulate eq.~\ref{eq:k_k'} as 
\begin{equation}  \label{eq:k_with_RT}
\small
    \begin{aligned}
     k' &= \frac{R_{off} + R_T + nr}{R_{on} + R_T + nr} = \frac{k-1}{1 + (R_T+nr)/R_{on}} + 1
    \end{aligned}
\end{equation}
$R_T$ is the resistance from a transistor. Figure~\ref{fig:Rt} shows $R_T$ values in different technology nodes. In 22\,nm and 28\,nm technology nodes $R_T \approx 1.7\,k\Omega$. $r \approx 2.5\,\Omega$, as estimated from the 4K ReRAM chip layout. 

\subsection{Scaling of Error Probability}
\label{sec:scaling_Perr}
\begin{figure}[!t]
\centering
\includegraphics[scale=0.35]{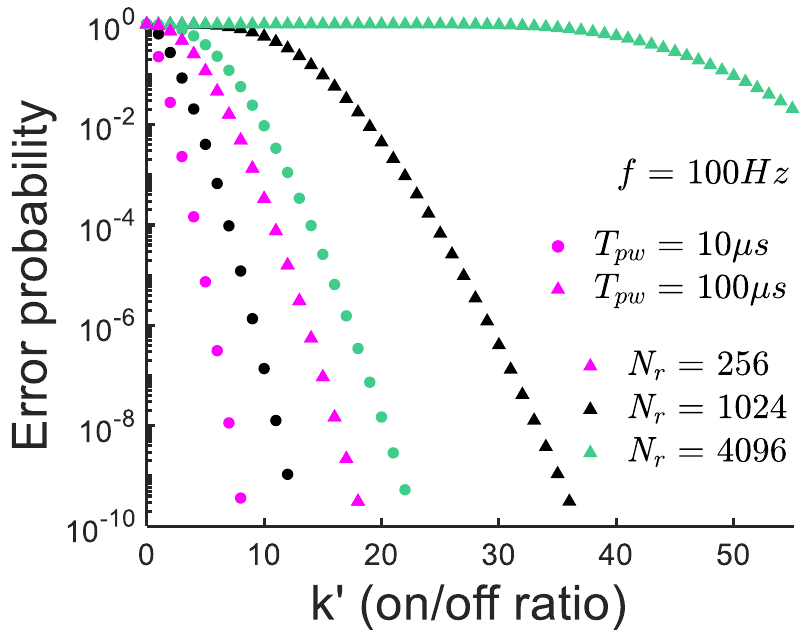}
\caption{Probability of an undesired error output pulse due to accumulated ``off" currents, with independent Poisson input pulse trains. $k'$ implies the number of simultaneous inputs that can be tolerated. $f$ is the frequency of each input, $T_{pw}$ is the routing pulse width, $N_r$ is the size of the crossbar.}
\label{fig:Perr}
\end{figure}

As the size of the router increases, the minimum on/off ratio required increases, due to higher probability of simultaneous inputs, as shown in Fig.~\ref{fig:Perr}. The theory of modeling the relationship between error probability, input frequencies, routing pulse width and the requirement of on/off ratio was presented in \cite{chen2022reliability}. The longer the pulse width, the higher the probability that more simultaneous pulses occur in $T_{pw}$, causing more memristor leakage current accumulation. 
For pulse widths ($T_{pw}$) between 100s\,ns to a few $\mu$s (e.g. neuromorphic circuits in sub-threshold~\cite{dynapse2017}), and for large  crossbar arrays ($N_r=4096$), the requirement of $k'$ is about 20, to achieve error probability $<10^{-10}$. For shorter $T_{pw}$ values of a few ns to tens of ns (e.g. above-threshold analog circuits in mixed-signal systems, or digital circuits), $k'=10$ can be used, reducing the error probability to $<< 10^{-10}$.

\begin{figure}[!t]
\centering
\includegraphics[scale=0.42, trim={0.15cm 0.1cm 0 0.cm},clip]{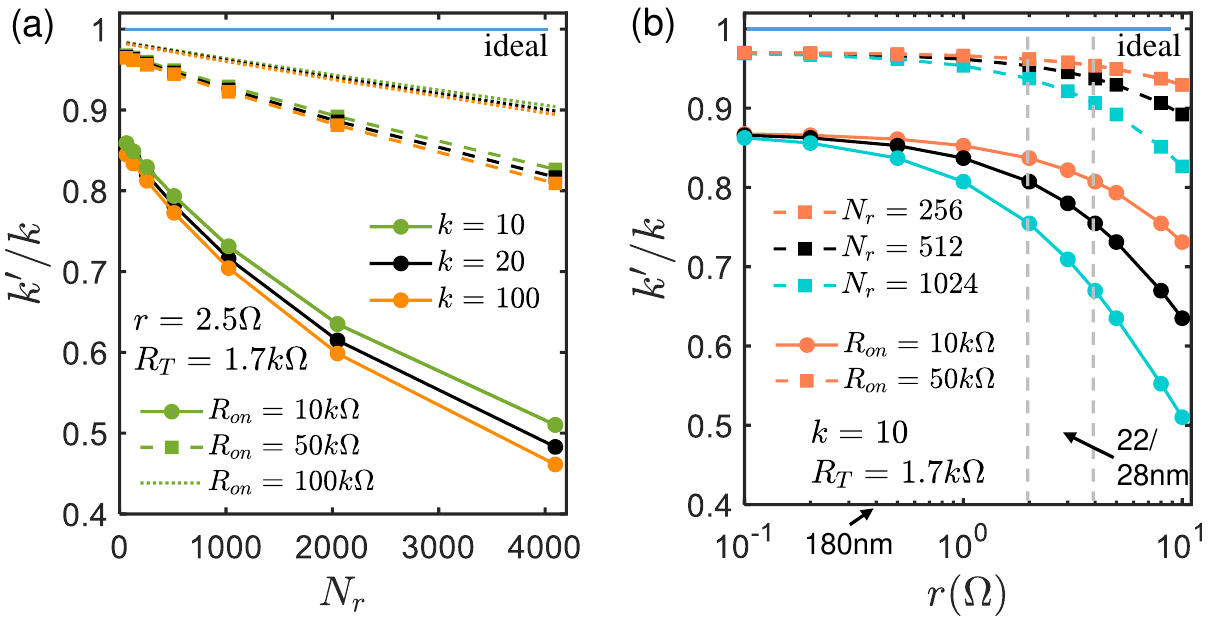}
\caption{The scaling of sensing margin with (a) router size $N_r$ and (b) line resistance $r$. $R_{on}$ is more dominant than $k$. When $N_r<512$, in 22/28\,nm technology nodes ($r\approx2.5\,\Omega$), the line resistance is negligible. }
\label{fig:kp_scaling}
\end{figure}

Figure~\ref{fig:kp_scaling} shows the scaling of sensing margin ($k'/k$) with router size and $r$, where
$R_{on}$ is more dominant than $k$. Increasing sensing margin at the cost of lowering $R_{on}$ is not an effective solution, because low $R_{on}$ ($<10\,k\Omega$) suffers serious IR drop. Instead, $R_{off}$ should increase. In 22/28\,nm technology, line resistance effect is negligible for $N_r<512$ when $R_{on}>10\,k\Omega$.

To support $1K$ inputs ($R_T\approx1.7\,k\Omega$, $nr\approx2.5\,k\Omega$), and to achieve relatively high $k'$ around 10, $R_{on}>10\,k\Omega$, $R_{off}>200\,k\Omega$ is required. In engineering devices with on/off ratio of approximately $20$, a resistance range of hundreds $k\Omega$ to $M\Omega$ is preferable (e.g. $R_{on} = 50\,k\Omega$, $R_{off} = 1\,M\Omega$). For mixed-signal neuromorphic chips that use circuits operating in the sub-threshold or weak-inversion regime, resistance ranges of $M\Omega$ for routing spikes are preferable. This takes advantage of the fact that these neuromorphic circuits typically have slow time constants, so the corresponding routers have less stringent speed margins and lower power requirements.

\subsection{Effect of Transistor Leakage Currents}
\label{sec:Rt_leak}

\begin{figure}[!t]
\centering
\includegraphics[scale=0.26, trim={0cm 0.82cm 0 0.1cm},clip]{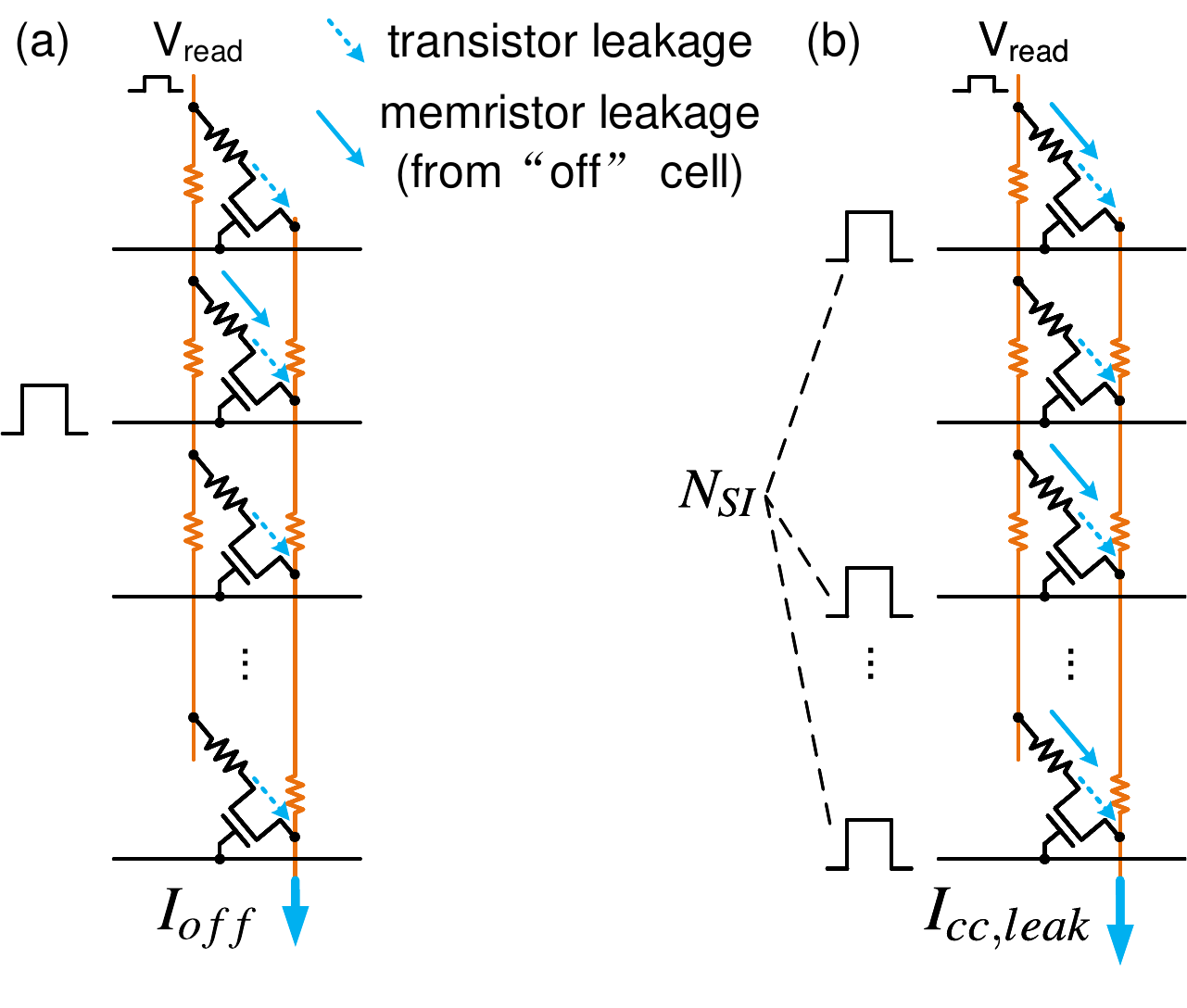}
\caption{Illustration of leakage current accumulation from transistors and ``off" memristors when input pulses arrive. (a) $I_{off}$ is the resulting current from one input at $R_{off}$. (b) $I_{cc,leak}$ is the overall ``off'' currents on a channel corresponding to the arrival of $N_{SI}$ simultaneous inputs.}
\label{fig:Iccleak_schematic}
\end{figure}

\begin{figure}[!t]
\centering
\includegraphics[scale=0.42, trim={0.15cm 0.05cm 0 0.1cm},clip]{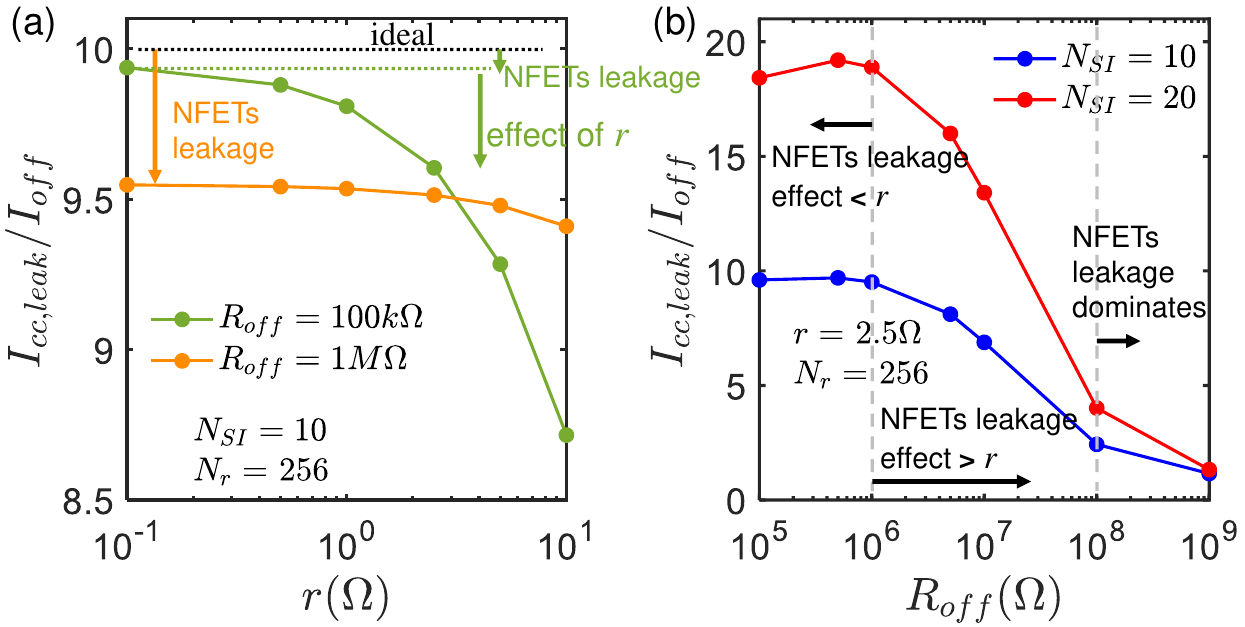}
\caption{The effect of line resistance and leakage currents from NFETs (22\,nm FDSOI) in 1T1R array when ``off'' cells receive simultaneous input pulses. (a) IR drop effect on the overall ``off'' currents $I_{cc,leak}$ of $N_{SI}$ simultaneous inputs. The lower the $I_{cc,leak}$, the lower the leakage power consumption and higher the routing reliability. (b) The influence of NFETs leakage vs. line resistance effect as $R_{off}$ increases.}

\label{fig:Icc_vs_Ioff}
\end{figure}

\begin{figure}[!t]
\centering
\includegraphics[scale=0.42, trim={0.15cm 0.05cm 0 0.15cm},clip]{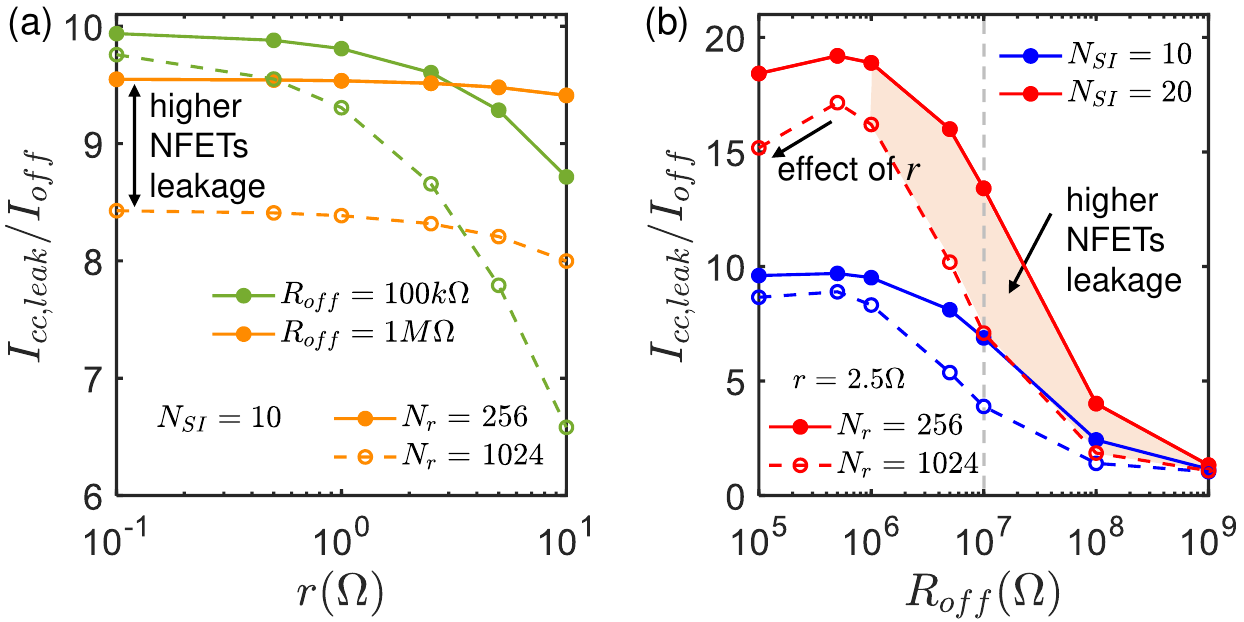}
\caption{Joint effects of IR drop and transistor leakage when the router size scales up from 256 to 1024.}
\label{fig:Icc_vs_Ioff_biggerNr}
\end{figure}

When including transistors and line resistance in SPICE simulation (using a normal threshold NFET of 22\,nm FDSOI technology), transistor leakage and IR drop play concurrent roles in overall leakage current accumulation when the router receives inputs. The corresponding schematic is shown in Fig.~\ref{fig:Iccleak_schematic}. The difference between $I_{cc,leak}/I_{off}$ and $N_{SI}$ indicates the influence by NFETs leakage and IR drop. 
The overall leakage current from 256 NFETs in one channel is about $10\,nA$. Leakage currents from NFETs have little effects when $R_{off}< 1\,M\Omega$ in Fig.~\ref{fig:Icc_vs_Ioff} (a), but they overpower currents from ``off'' cells when $R_{off}>100\,M\Omega$ in Fig.~\ref{fig:Icc_vs_Ioff} (b). 
Figure~\ref{fig:Icc_vs_Ioff} (a) also shows that the accumulated currents from $N_{SI}$ ``off'' cells ($I_{cc,leak}$) in a channel is lower than $N_{SI} \times I_{off}$ (ideal) due to IR drop, which slightly enlarge the sensing window of distinguishing from $R_{on}$ compared to the ideal case. In other words, IR drop lowers $I_{cc,leak}$ and alleviates reliability issue regarding accumulation of $I_{off}$. But this alleviation is trivial when $R_{off}>1\,M\Omega$.

As the router size scales up from 256 to 1024, as illustrated in Fig.~\ref{fig:Icc_vs_Ioff_biggerNr}, the joint effects of NFETs leakage and IR drop undergo notable changes. NFETs leakage increases fourfold, concurrently with an increase in IR drop. IR drop can be disregarded at $r = 0.1\,\Omega$, so the shift from $N_r = 256$ to 1024 primarily reflects the impact of the elevated overall NFETs leakage current, as depicted in Fig.~\ref{fig:Icc_vs_Ioff_biggerNr} (a). It also shows that for $R_{off} > 1\,M\Omega $, IR drop becomes negligible. Thus, within the range of $R_{off} > 1\,M\Omega$, the changes in Fig.~\ref{fig:Icc_vs_Ioff_biggerNr} (b) due to the scaling up of router size is solely attributable to the increase in overall NFETs leakage currents.
Furthermore, $I_{cc,leak}/I_{off}$ is decreased by over 50\% compared to the ideal case when $R_{off}>10\,M\Omega$ (e.g., $I_{cc,leak}/I_{off} < 5$ at $N_r=1024$, $N_{SI}=10$). Consequently, NFETs leakage becomes predominant when $R_{off}>10\,M\Omega$.

In addition, the leakage currents indicate a design trade-off for 1T1R arrays between area and leakage power consumption from transistors in 22\,nm technology and beyond: if using high $V_T$ FETs for lower leakage, bigger size is required to achieve higher driving ability for memristor programming.

\section*{Conclusions}
We determined specifications for enabling the scaling of memristor crossbar routers with theoretical considerations and simulation studies, and validated them with experimental measurements performed on a 4K-ReRAM chip. We configured the chip to implement a crossbar router and demonstrated experimentally its limits in transmitting spikes for SNN neuromorphic architectures.
The routing error probability caused by current accumulation from ``off'' cells and the effect of the IR drop on metal lines is evaluated. We showed how the IR drop decreases the read margin (on/off ratio), and increases the error probability.
We proposed an analytic model that can be used to guide future development of memristive devices for these use cases.
We concluded that, for the examples at 22\,nm technology node, a resistance range of LRS $>50\,k\Omega$ with on/off ratio $>20$ is ideal. Our study also indicates a design trade-off between area and leakage power consumption.

{
\bibliographystyle{IEEEtran}
\bibliography{reference}
}

\end{document}